\documentclass[11pt,tightenlines,eqsecnum,floats,aps,amsmath,amssymb,nofootinbib,prd,shownopacs]{revtex4}
\usepackage{amsmath,amssymb,amsfonts,amsthm}
\usepackage{graphicx}
\usepackage{enumerate} % advanced enumerate environment
\usepackage{colordvi} % for color text
\usepackage[mathscr]{eucal}
\usepackage[latin1]{inputenc}
\usepackage{amsmath}
\usepackage{amsfonts}
\usepackage{amssymb}
\usepackage{graphicx}

\newcommand{\rcr}{\rho_{\mathrm{crit}}}

\newcommand{\f}{\frac}

\def\rcr{\rmax}

\usepackage{enumerate}

\def\f{\frac}

\def\d{\textrm{d}}

\usepackage{enumerate}

\newcommand{\be}{\nopagebreak[3]\begin{equation}}
\newcommand{\ee}{\end{equation}}
\newcommand{\ba}{\nopagebreak[3]\begin{eqnarray}}
\newcommand{\ea}{\end{eqnarray}}

\newcommand{\bmult}{\nopagebreak[3]\begin{multline}}
\newcommand{\emult}{\end{multline}}

\def\d{{\rm d}}

\def\f{\frac}

\def\d{\textrm{d}}

\def\d{{\rm d}}
\def\rcr{\rho_{c}}

\allowdisplaybreaks

\usepackage{enumerate}
\begin{document}

\title{On the relationship between the modifications to the Raychaudhuri equation and the canonical Hamiltonian structures}

\author{Parampreet Singh$^{*}$}
\author{S. K. Soni$^\dagger$}
\affiliation{${}^*$Department of Physics and Astronomy,
Louisiana State University, Baton Rouge, LA 70803, U.S.A. \\
%\vskip0.05cm
\,
\,
\,
\\
${}^\dagger$Department of Physics, Sri Guru Tegh Bahadur Khalsa College, \\ University of Delhi, Delhi 110007, India}
%\,
%\,
%\,

\begin{abstract}
The problem of obtaining canonical Hamiltonian structures from the equations of motion, without any knowledge of the action, is studied in the context of the spatially 
flat 
Friedmann-Robertson-Walker models. Modifications to Raychaudhuri equation are implemented 
independently as quadratic and cubic terms of 
energy density  without introducing additional degrees of freedom. Depending on their sign, modifications make gravity 
repulsive above a curvature scale for matter satisfying strong energy condition, or more attractive than in the classical theory. Canonical structure of the 
modified theories is determined 
demanding that the total Hamiltonian be a linear combination of gravity and matter Hamiltonians. 
 In the quadratic repulsive 
case, the modified canonical phase space of gravity 
is a polymerized phase space with canonical momentum as inverse trigonometric function of Hubble rate; the canonical Hamiltonian can be 
identified with the effective Hamiltonian in loop quantum cosmology. The repulsive cubic modification results in a 
`generalized polymerized' canonical phase space. Both of the repulsive modifications are found to yield singularity avoidance.
In contrast, the quadratic and cubic attractive modifications result in a canonical phase space in which canonical momentum is 
non-trigonometric and singularities persist.  
Our results hint on connections between repulsive/attractive nature of modifications to gravity arising from 
gravitational sector and polymerized/non-polymerized gravitational phase space.

\end{abstract}

\maketitle

\section{Introduction}

Suppose one desires a particular form of the dynamical equations in a modified theory of gravity, and has no knowledge of the Lagrangian or the Hamiltonian 
structure of the theory. Then, what properties of the 
canonical Hamiltonian structure can one deduce directly from dynamical equations? What does a cosmological dynamical equation in a modified theory of gravity tell 
us about the underlying canonical phase space? From an 
inverse point of view, these are fundamental questions which can help in deciphering the right Hamiltonian (and the action) for a theory 
which yields desired physics. As an example, one may demand that there be a modified theory of gravity free of singularities without violating 
weak energy condition. Starting from the modified dynamical equations in general relativity (GR), which avoid singularities for positive definite energy density, how do we  
systematically derive the corresponding canonical Hamiltonian without any knowledge of the action? 

The goal of this manuscript is to address these questions for spatially flat homogeneous and isotropic models in modified gravity scenarios. Due to homogeneity, the diffeomorphism constraint is 
trivially satisfied and the only non-trivial constraint is the Hamiltonian 
constraint which vanishes weakly. The modified cosmological theories we consider are assumed to have same degrees of freedom as 
classical cosmology without any violation of  
conservation law of matter-energy, and, general covariance.  The modification to 
cosmological dynamics is assumed to arise purely from 
the gravitational sector. Matter is assumed to be coupled minimally. We focus our attention on the conditions imposed by the properties of 
gravity, which make it repulsive above a curvature scale without violation of energy conditions or more attractive than GR, on the possible forms of the underlying 
canonical Hamiltonian structures  producing them. Our discussion will be based on the existence of the total Hamiltonian as a linear combination ${\cal H} = {\cal H}_g + {\cal H}_m$ of 
gravity and matter Hamiltonians. To find the Hamiltonian in this form we rely on the fact that the matter Hamiltonian is given by ${\cal H}_m = \rho V$, where $\rho$ is the energy density of matter and $V$ is the cube of the scale factor $(a)$, and that in the modified gravity scenarios considered in our analysis matter Hamiltonian is not affected.

The problem we wish to solve has parallels in classical mechanics.
It is of great importance in mechanics to find the procedure of systematically arriving at the Lagrangian and the Hamiltonian of a 
system starting from 
its equations of motion (see for eg. \cite{santilli,sarlet,jan,jmp,desloge,berger,huang,soni,currie,hojman2,hojman}). Conventionally, one guesses a form for the 
Lagrangian from deep physical intuition  (and obtains the corresponding Hamiltonian), and then 
demonstrates the correctness of the Lagrangian (or the Hamiltonian) by verifying that the resultant Lagrange's equation of motion (or the 
Hamilton's equations) are indeed the 
known equations of motion. However, starting from equations of motion different strategies exist to find Lagrangian and Hamiltonian. An 
action can be obtained using Helmholtz conditions in the the inverse problem of calculus of variations from which a 
Hamiltonian can be derived \cite{santilli}. A Hamiltonian can also be obtained from equations of motion bypassing the Lagrangian 
 using constants of motion and symmetry vectors \cite{currie,hojman2,hojman}. For example, in the case of the 
electromagnetic field, the form of Lagrangian density is conventionally guessed from covariance.  Verification of this guess is 
demonstrated by showing that the Lagrange's equations yield the Maxwell equations. However, the 
Lagrangian density can be systematically derived from Maxwell's equations 
using reverse procedure based on Hamilton's principle \cite{huang}. As another example, an action principle for extended objects with arbitrary multipole moments  in general relativity  can be obtained by expanding stress energy tensor in a Taylor series and suitable contractions of resulting terms with spacetime curvature components \cite{jeeva}. Whereas, similar action upto the dipole moment has been obtained using inverse variational procedure for a spinning particle \cite{derz}. Similarly, non-Lagrangian construction of Hamiltonian structures has been performed in a variety of cases, such as for 
Korteweg-de Vries equation using a symmetry vector and a constant of the motion \cite{hojman2}, and, for motion of projectiles with resistance 
using one constant of the motion \cite{hojman}.

Our approach for the discussion of the problem posed in this work is inspired by an analog of second order equation of motion in the cosmological models. This is the 
Raychaudhuri equation for the scale factor of the universe. For matter with a given equation of state, the scale factor $a$ and its time 
derivative encode all dynamics which can be obtained from Raychaudhuri equation. This equation plays an important role in gravitational 
theories. It is central in understanding geodesic flows 
and singularity theorems, and as  a `force law' it reveals  
attractive ($\ddot a < 0$) or repulsive ($\ddot a > 0$) nature of gravity. 
Given the classical Raychaudhuri equation or one of its avatars in modified gravity, 
the sign of $\ddot a$ can be changed by choosing appropriate matter. For example, in classical cosmology for matter which obeys 
(or violates) strong energy condition, the classical Raychaudhuri equation yields $\ddot a < 0$ (or $\ddot a > 0$). However, in general 
for modified gravity scenarios the repulsive/more attractive nature of gravity deduced from Raychaudhuri equation is not  restricted to the 
choice of matter. Due to an interplay of gravitational and matter sectors, Raychaudhuri equation in a modified gravity scenario 
may take a form such that the sign of $\ddot a$ can become positive even for matter which does not violate strong energy condition, and 
vice versa. It is in this latter sense, we discuss the attractive/repulsive property of modified gravity in our analysis. In particular, a modification of the 
Raychaudhuri equation will be denoted repulsive (attractive) if it yields $\ddot a > 0$ ($\ddot a < 0$) for matter for which gravity is attractive in the classical theory.

We provide a straightforward analytic procedure to find the Hamiltonian in terms of canonical variables that requires no physical intuition 
other than the realization that it is useful to look for a constant of the motion ($C$) in dynamics encoded by the Raychaudhuri equation. This 
constant of the motion can then be identified with an intermediate Hamiltonian using Hojman's analysis \cite{hojman}. However, we show that the latter step is not sufficient to 
understand the Hamiltonian structure, in particular of the gravitational sector due to the following reason. The intermediate Hamiltonian does not appear as a 
linear combination of gravity and matter Hamiltonians. The matter Hamiltonian, which is simply related to the 
energy density as ${\cal H}_m = \rho a^3$, appears in product with a function of phase space or even with higher powers in modified 
 scenarios. For example, in classical cosmology the intermediate Hamiltonian 
turns out to be of the form $C(a,p) = p^2/2m + V(a)$, where $p = \dot a$ is the conjugate momentum and $V(a)$ is a function of scale factor. The potential term does not correspond to the matter Hamiltonian. Since this term is this sense mixed, the ``kinetic energy'' term can not be identified as the gravitational Hamiltonian. Rather in the modified 
gravity scenarios, this mixing gets more complicated and the potential term becomes a non-linear function of matter Hamiltonian and phase space variables. This obstacle is overcome  in our procedure by using the property that the total Hamiltonian vanishes. We are then able to write a Hamiltonian in the form ${\cal H}_g + {\cal H}_m$ and identify the canonical phase space structure. 

We consider four different types of modified Raychaudhuri equations and derive the resulting canonical phase space structure and Hamiltonians. The modifications to the classical Raychaudhuri equation involve  $\rho^2$ and $\rho^3$ terms, with positive as well as negative signs. %Both the signs of the modifications are considered. 
For the positive sign modifications, $\ddot a$ changes sign above a certain energy density $\rho_c$. The scale $\rho_c$ is a parameter of the modified Raychaudhuri equation supposed to be fixed by the underlying theory. If one starts with matter which leads to attractive gravity in GR, modified Raychaudhuri equations with $+ \rho^2$ and $+ \rho^3$ modifications, result in repulsive gravity for $\rho > \rho_c$. For this reason we will call these modifications as repulsive quadratic and repulsive cubic modifications respectively. The quadratic and cubic modifications with negative sign result in making gravity universally more attractive than in GR for matter which results in classical attractive gravity. We label these modifications as attractive quadratic and attractive cubic modifications. For the repulsive modifications, we find the Hubble rate to be universally bounded which signal 
a generic resolution of cosmological singularities \cite{ps09}. For the quadratic repulsive case, we find the resulting canonical phase space to be polymerized. The canonically conjugate momentum to volume, which we choose as the generalized coordinate in all the cases, turns out to be an inverse trigonometric function of Hubble rate. The canonical Hamiltonian can be identified with the effective Hamiltonian in loop quantum cosmology with an appropriate identification of energy density scale at which gravity becomes repulsive \cite{ps06,aps3}. In the cubic repulsive case, 
canonical momentum is a hypergeometric function of the inverse trigonometric function of the Hubble rate. With similarities to the polymerized momentum, we 
label this momentum as a `generalized polymerized' version of the one obtained in the quadratic repulsive case. Unlike the quadratic repulsive case whose cosmological dynamics has been extremely well studied in the framework of loop quantum cosmology  \cite{as}, the cubic repulsive case is a new non-singular cosmological model. In contrast to the repulsive modifications,  attractive modifications do not yield conjugate momentum which is an inverse trigonometric function and thus there is no polymerization in the phase space. Hubble rate is unbounded in these cases. Thus, there is a sharp distinction between the 
gravitational phase spaces of repulsive and attractive modifications.% considered in our analysis.

The structure of this manuscript is as follows. In Sec. II, we  illustrate our method to obtain a canonical Hamiltonian as a linear combination of gravitational and matter Hamiltonians directly from the  
Raychaudhuri equation for the case of classical cosmology. This example serves as a template for more general examples considered 
in the subsequent sections with some additional steps. Without any loss of generality, we consider the Raychaudhuri equation for the case of dust which has constant matter Hamiltonian. It may be noted that choosing a different matter with a fixed equation of state leads to no change in procedure. It turns out that the resulting Hamiltonian yields evolution for matter for any equation of state by replacing the matter Hamiltonian accordingly. 
Following the procedure laid down in Sec. II, 
the canonical Hamiltonian structure for a $\rho^2$ modification with a positive sign in the classical Raychaudhuri equation is obtained in Sec. III. 
In Sec. IV, we discuss the derivation of the canonical Hamiltonian structure for $\rho^2$ modification with a negative sign in the classical Raychaudhuri equation. 
To find the canonical Hamiltonian in the form ${\cal H}_g + {\cal H}_m$, quadratic cases require obtaining roots of the vanishing of the intermediate Hamiltonian which is a quadratic equation in 
energy density. Cubic modification in energy density with a positive sign to the classical Raychaudhuri equation is analyzed in Sec. V and its counterpart with a negative sign is discussed in Sec. VI. In these cases, to obtain canonical phase space one finds roots of a roots of  a cubic equation in $\rho$. Physical roots are determined demanding energy density is real and positive. The latter requirement is needed only for the attractive modifications.  Both of the repulsive cases, lead to dual roots covering different sectors of the gravitational phase space. The attractive modifications yield a single root which covers the entire range of energy density. We keep the conventions to discuss the phase space structures for different Hamiltonians same in  Sec. III-VI  (which follow the convention set in Sec. II).  
The manuscript concludes with a discussion in Sec. VII.

\section{Classical Hamiltonian from the Raychaudhuri equation}
In this section, we illustrate the procedure of obtaining the Hamiltonian in terms of canonical variables from the Raychaudhuri 
equation in the spatially flat FRW universe in classical GR. Key elements of the method we outline here, are used in 
subsequent sections for modified gravity scenarios.   The Raychaudhuri equation 
for a perfect fluid with energy density $\rho$ and pressure $P$, in classical FRW model is given by:
\be\label{raiclassical}
\ddot a = - \f{4 \pi G}{3} (1 + 3 w) \rho \, a ~.
\ee
Here $w$ is the equation of state $w = P/\rho$, and the `dot' is the derivative with respect to proper time. The energy density is defined as $\rho = {\cal H}_m/V$ where ${\cal H}_m$ is the matter Hamiltonian and $V$, equal to $a^3$, denotes the volume of the universe. Pressure is defined as $P = -\partial {\cal H}_m/\partial V$. The energy density and pressure satisfy the conservation law following from the covariant conservation of the stress-energy tensor:% 
\be \label{cons}
\dot \rho + 3 H (\rho + P) = 0 ~,
\ee
where $H = \dot a/a$ is the Hubble rate.

For simplicity, let us consider the case of equation of state $w=0$. In this case, the Raychaudhuri equation can be rewritten as 
\be\label{raidust}
\ddot a = - \f{4 \pi G}{3} \rho \, a ~.
\ee
Then, using (\ref{cons}) one finds that $\rho = \rho_o (a_o/a)^3$, with $\rho_o$ and $a_o$ being constants of integration. 
Thus, for the case of dust the matter Hamiltonian is a constant ${\cal H}_m = \rho a^3 = \rho_o a_o^3$.

We are interested in finding the Hamiltonian ${\cal H}$ corresponding to eq.(\ref{raidust}) in terms of the canonical phase space variables as a linear combination ${\cal H} = {\cal H}_g + {\cal H}_m$. The first step is to find a constant of the motion starting from the Raychaudhuri equation. This constant of the motion can be identified as a Hamiltonian \cite{hojman}. However, this Hamiltonian serves only as a intermediate Hamiltonian in our analysis.  As we will see, the intermediate Hamiltonian is not in the form of a linear combination of gravitational and matter Hamiltonians. The canonical Hamiltonian in the desired form  is obtained using the property that the (intermediate) Hamiltonian vanishes due to general covariance.

To set the stage, let us consider a system described by a second order dynamical equation 
\be\label{fg}
\ddot q = A(q) B(\dot q) ~,
\ee
where $q$ denotes a generalized coordinate. The second order equation can be written as two 
 first order equations as
\be\label{x1x2hojman}
\dot x_1 = x_2, ~~~~~~~~~ \dot x_2 = A(x_1) B(x_2) ~,
\ee
where $x_1$ and $x_2$ are defined as $x_1 = q$ and $x_2 = \dot q$ respectively. It is straightforward to see that eq.(\ref{fg}) yields the following time independent constant:
\be \label{Constant}
C(q,\dot q) = - \int A(q) \d q \, + \, \int  B^{-1}(\dot q) \, \dot q \, \d \dot q ~. \\
\ee

Once a constant of the motion is available, one uses Hojman's analysis, where it is shown that Hamiltonian structure corresponding to the dynamical equation (\ref{fg}) is obtained by identifying 
$C(x_1, x_2)$ as a Hamiltonian with $x_1$ and $x_2$ as the phase space variables \cite{hojman}. These variables satisfy the Poisson bracket $\{x_1,x_2\} = \mu(x_1,x_2)$. The function $\mu(x_1,x_2)$, denoted as $\mu$ in the following, is to be determined from the consistency of the Hamiltonian evolution using the first order equations of motion (\ref{x1x2hojman}). It turns out that 
$\mu = B(x_2)$.

Above steps, i.e. finding a constant of the motion and treating that constant as a Hamiltonian, can be adapted to cosmological scenarios. However, the resulting Hamiltonian $C(x_1,x_2)$ does not turn out to be a  linear combination of gravitational and matter 
Hamiltonians. To obtain such a Hamiltonian we use the fact  that total Hamiltonian vanishes. 
We now apply our method to the classical Raychaudhuri equation (\ref{raidust}). This equation is already in the form (\ref{fg}), with $q$ identified as the scale factor. To express the classical Raychaudhuri equation (\ref{raidust}) in terms of two first order equations, we choose
\be\label{x1x2def}
x_1 = a, ~~~~ \mathrm{and} ~~~~ x_2 = \dot a ~.
\ee
Thus, eq.(\ref{raidust}) can be written as
\be \label{classx1x2}
\dot x_1 = x_2, ~~~~~ \dot x_2 = -\f{4 \pi G}{3} \rho x_1 ~.
\ee
Comparing with (\ref{x1x2hojman}), we obtain,
%It can be expressed  we get
\be
A(a) = -\f{4 \pi G}{3} \rho, ~~~~ \mathrm{and} ~~~~ B(\dot a) = 1 ~.
\ee

Substituting $A(a)$ and $B(\dot a)$ in (\ref{Constant}), and using $\rho = {\cal H}_m/x_1^3$ we obtain a constant of the motion 
in terms of $x_1$ and $x_2$:
\be\label{ham1}
C(x_1, x_2) = \f{x_2^2}{2} - \f{4 \pi G}{3} {\cal H}_m  x_1^{-1} ~.
\ee

\vskip0.5cm

\noindent
It is easy to check that $C(x_1,x_2)$ leads to a consistent Hamiltonian evolution by computing the Hamilton's equations: 
\be
\dot x_1 = \mu \f{\partial C}{\partial x_2}  ~~~ \mathrm{and} ~~~~ \dot x_2 = - \mu \f{\partial C}{\partial x_1} ~.
\ee
These equations yield, 
\be
\dot x_1 = \mu x_2 ~~~ \mathrm{and} ~~~ \dot x_2 = - \f{4 \pi G}{3} {\cal H}_m \mu x_1^{-2} ~.
\ee
A comparison of the above equations with (\ref{classx1x2}), shows that $\mu = 1$ for the Hamiltonian evolution generated by $C(x_1,x_2)$.  Hence, $x_1$ and $x_2$ are canonically conjugate phase space variables for the Hamiltonian in eq.(\ref{ham1}). 
The phase space variables $x_1$ and $x_2$ transform as scalars under time reparameterization and $C(x_1,x_2)$ vanishes weakly due to general covariance \cite{teitelboim}.

Although $C(x_1,x_2)$ yields a consistent Hamiltonian evolution, it is not in the form ${\cal H}_g + {\cal H}_m$. To obtain a Hamiltonian in this form from $C(x_1,x_2)$ we note that for 
a function on phase space $f(x_1,x_2)$ which is non-divergent and has non-divergent derivatives with respect to $x_1$ and $x_2$, 
$f(x_1,x_2) C(x_1,x_2)$ is also a constant of the motion.\footnote{The function $f(x_1,x_2)$ can be interpreted as  a different choice of the lapse function.} This is straightforward to see by using the property that $C(x_1,x_2)$ is a constant, 
 and  $C(x_1,x_2) \approx 0$. To obtain a Hamiltonian in the desired form we 
 we multiply $C(x_1,x_2)$ by $(-3/4 \pi G) x_1$, and obtain a Hamiltonian 
\be \label{ham2}
{\cal H} = -\f{3}{8 \pi G} x_2^2 x_1 + {\cal H}_m 
\ee
which weakly vanishes. 
This Hamiltonian yields the same space of solutions as $C(x_1,x_2)$ and is in the form ${\cal H}_g + {\cal H}_m$, with ${\cal H}_g$ identified as the term not containing ${\cal H}_m$ in the above equation.

Before we investigate the canonical structure of this Hamiltonian, it is interesting to note that using $C(x_1,x_2) \approx 0$, ${\cal H}$ can be written as %the  Hamiltonian constraint can  be written as 
\be \label{classicalHamrel}
{\cal H} = - \rho a^3 + {\cal H}_m \approx 0 ~,
\ee
The expression ${\cal H}_m - \rho a^3$ is a constant of the motion, which also follows independently from the matter-energy conservation law.

Having obtained the Hamiltonian in the desired form, we now find the canonical phase variables. It turns out that $x_1 = a$ and $x_2 = \dot a$ are non-canonical pair under the Hamiltonian flow generated by ${\cal H}$. 
The Hamilton's equations corresponding to ${\cal H}$ in eq.(\ref{ham2})  require 
$\{x_1, x_2\} = \mu = - (4 \pi G/3) x_1^{-1}$. The canonical momentum corresponding to $x_1 = a$ can then be found by 
\be
p_a \, = \, \int \mu^{-1} \d x_2 \, = \, -\f{3}{4 \pi G} x_1 x_2 ~.
\ee

In terms of the canonical variables $(a,p_a)$, the Hamiltonian constraint can be written as 

\be \label{hamcon2}
{\cal H}(a,p_a) = -\f{2 \pi G}{3} p_a^2 a^{-1} + {\cal H}_m \approx 0 ~.
\ee
This gives the Hamiltonian for the spatially flat FRW model in classical general relativity for matter specified by ${\cal H}_m$. %This corresponds to the ADM Hamiltonian for the spatially flat homogeneous and isotropic model. 
An alternative way to write this Hamiltonian is by choosing volume as the generalized coordinate, i.e. $x_1 = V$,
whose conjugate momentum is proportional to the Hubble rate: $p_V = - (4 \pi G)^{-1} H$. In terms of canonical variables $(V, p_V)$, we obtain the Hamiltonian constraint as:
\be \label{hamcon3}
{\cal H}(V,p_V) = -\, 6 \pi G p_V^2 V + {\cal H}_m \approx 0 ~.
\ee
 Thus, we obtain a Hamiltonian for the classical spatially flat FRW universe in the 
form ${\cal H}_g + {\cal H}_m$ in the canonical phase 
space variables $a$ and $p_a$, as well as $V$ and $p_V$. In the following sections, we  use $V$ as a generalized coordinate in the final Hamiltonian, and will suppress the volume subscript in its canonical momentum.

It is important to note that even though we 
started with dust as the matter content, the above Hamiltonian provides consistent Hamiltonian evolution for arbitrary matter content in 
classical cosmology. One could have started this computation for any matter with a fixed equation of state and reach the same canonical Hamiltonian. To obtain dynamics of any given matter with a fixed equation of state, one has to choose the corresponding matter Hamiltonian in eq.(\ref{hamcon3}), or in any of the equivalent Hamiltonians derived above. It is easily seen that the vanishing of ${\cal H}$ yields the classical Friedmann equation:
\be
H^2 = \frac{8 \pi G}{3} \frac{{\cal H}_m}{V} ~.
\ee
It is known from the classical dynamics that the physical solutions resulting from the Hamiltonian in eq.(\ref{hamcon3}) are singular when weak energy condition is satisfied.

\section{The Raychaudhuri equation with a quadratic repulsive modification in energy density}
In this section, we consider the first of our modifications to the classical Raychaudhuri equation and derive the corresponding canonical 
Hamiltonian structure.  As before, we consider the case for dust and introduce a $\rho^2$ modification to eq.(\ref{raidust}) as:  
\be \label{modrai}
\ddot a = - \f{4 \pi G}{3} \rho \left(1 - \f{\rho}{\rcr} \right) \, a ~.
\ee
Here $\rcr$ is a constant energy density whose value is to be determined by the underlying theory which is supposed to lead to the above 
equation.  The modification  is introduced without affecting the underlying degrees of freedom of the gravitational and matter sectors. As is 
the classical theory, the modified theory of gravity is assumed to be generally covariant, whose total Hamiltonian ${\cal H}$ weakly vanishes. 
The energy density $\rho$ satisfies the conservation law (\ref{cons}). 
At energy densities $\rho \ll \rcr$, eq.(\ref{modrai}) is approximated by the Raychaudhuri equation in the classical theory (\ref{raidust}). 
When the energy density becomes greater than $\rcr$, the acceleration term in the Raychaudhuri equation changes sign. Gravity becomes 
repulsive at these scales. Note that the repulsive nature of gravity occurs without changing the equation of state of matter. 

Our goal is to find the Hamiltonian corresponding to the above modified Raychaudhuri equation in the form 
${\cal H} = {\cal H}_{g} + {\cal H}_m$. For dust, the matter Hamiltonian ${\cal H}_m =  \rho_o a_o^3$ is a constant. To find the gravitational part of the Hamiltonian ${\cal H}_g$ we employ the method outlined in the 
previous section for the classical Raychaudhuri equation. We start with choosing the gravitational phase space variables as in the 
classical case (eq.(\ref{x1x2def})):
\be%\label{x1ax2dota}
x_1 = \nonumber a ~~~~~~~~ \mathrm{and} ~~~~~~~~ x_2 = \dot a ~.
\ee
In terms of $x_1$ and $x_2$, the modified Raychaudhuri equation (\ref{modrai}) results in,
\be
\dot x_1 = x_2, ~~~ \dot x_2 = -\f{4 \pi G}{3} \rho \left(1 -  \f{\rho}{\rcr}\right) x_1 ~.
\ee
Comparing the above set with eqs.(\ref{x1x2hojman}), we identify:
\be
A(x_1) = -\f{4 \pi G}{3} \rho \left(1 - \f{\rho}{\rcr}\right) x_1, ~~~~ \mathrm{and} ~~~~ B(x_2) = 1 ~.
\ee
The constant of the motion $C(x_1,x_2)$ can then be determined using eq.(\ref{Constant}). It turns out to be,
\be \label{hojmanH}
C(x_1,x_2) = \f{x_2^2}{2} - \f{4 \pi G}{3} \f{{\cal H}_m}{x_1^3} \left(1 - \f{1}{4 \rcr} \f{{\cal H}_m}{x_1^3} \right) x_1^2
\ee
where we have used $\rho = {\cal H}_m/a^3$. It is straightforward to verify that this constant of the motion serves as a Hamiltonian for the modified Raychaudhuri 
equation (\ref{modrai}). However, it is not in the form ${\cal H}_g + {\cal H}_m$.  Its vanishing, i.e. $C(x_1,x_2) \approx 0$, yields the 
physical solutions. These solutions are restricted to those with energy density $\rho \leq 4 \rcr$.    

In contrast with the expression in the 
classical theory, $C(x_1,x_2)$ is not linear in ${\cal H}_m$. 
To write eq.(\ref{hojmanH}) in the form linear in matter Hamiltonian, we solve for the roots of ${\cal H}_m/a^3$ using $C(x_1,x_2) \approx 0$. It results in the  quadratic equation:
\be\label{quadx}
\xi (1 - \xi) = \f{3}{32 \pi G} \f{H^2}{\rcr}
\ee
where $\xi$ is defined as $\xi := \rho/4 \rcr$. This quadratic equation admits following roots:
\be
\xi_\pm = \f{1}{2} (1 \pm \sqrt{1 - 4 \alpha^2 H^2}), ~~~~~~%~~~~ \mathrm{and} ~~~~ \xi_- = \f{1}{2} (1 - \sqrt{1 - 4 \alpha^2 H^2}), ~~~ 
\mathrm{with} ~~~~ \alpha^2 := \f{3}{32 \pi G \rcr} ~.
\ee
Note that these roots individually do not span the whole range of energy density. For the negative root ($\xi_-$), energy density is in the range $0 \leq \rho \leq 2 \rcr$.  The positive root ($\xi_+$)  corresponds to the range $2 \rcr \leq \rho \leq 4 \rcr$. Both the independent roots are therefore essential to capture the complete dynamics in the phase space. 
In this modified gravity scenario, the Hubble rate is bounded above with a maximum value $|H| = 1/2 \alpha$ satisfied at $\xi_+ = \xi_- = 
1/2$. It vanishes at $\xi_- = 0$ and $\xi_+ = 1$. The former value corresponding to the regime where the energy density vanishes, and the latter 
where $\rho$ takes its maximum allowed value $\rho = 4 \rho_c$. From the modified Raychaudhuri equation (\ref{modrai}), one finds that the 
scale factor bounces at $\xi_+ = 1$. Unlike the classical theory, the physical solutions are non-singular.\\

The vanishing of $C(x_1,x_2)$ results in the Hamiltonian constraint in the desired form ${\cal H}_g + {\cal H}_m$
\be \label{quadrepeq}
{\cal H} = - 4 \rcr V \xi_\pm + {\cal H}_m \approx 0 ~
\ee
which is the analog of eq.(\ref{classicalHamrel}) for the quadratic repulsive modification.
%which is a  
We thus obtain two equations for the roots $\xi_+$ and $\xi_-$: 
\be\label{quadH-}
%{\cal H}^+ := - 
-\f{3 V}{8 \pi G \alpha^2}\,  \f{1}{2} (1 - \sqrt{1 - 4 \alpha^2 H^2}) + {\cal H}_m \approx 0 ~ ~~~\mathrm{for} ~ ~~0 \leq \rho  \leq 2 \rcr
\ee
and 
\be\label{H+}
%{\cal H}^- := 
- \f{3 V}{8 \pi G \alpha^2}\,  \f{1}{2} (1 + \sqrt{1 - 4 \alpha^2 H^2}) + {\cal H}_m \approx 0 ~ ~~~\mathrm{for} ~ ~~2 \rcr \leq \rho  \leq 4 \rcr
\ee
The L.H.S. of the above constraints is a constant of the motion, and provide us Hamiltonians in the respective ranges of energy density. 
We denote the Hamiltonians corresponding to negative and positive roots as $ {\cal H}^-$ and ${\cal H}^+$ respectively.

We now proceed to express ${\cal H}^-$  and ${\cal H}^+$ in terms of the canonical phase space variables. Given that the Hamiltonians ${\cal H}^+$ and ${\cal H}^-$ are explicit functions of volume $V$ and Hubble rate $H$, it is convenient to choose these as the gravitational phase space variables:
\be
x_1 = V, ~~~~ \mathrm{and} ~~~~ x_2 = H
\ee
where $x_1$ and $x_2$ satisfy the Poisson bracket:
\be
\{x_1,x_2\}^{\pm} = \mu^\pm ~.
\ee
Here $\pm$ in the above Poisson bracket relation imply that $+ve$ and $-ve$ roots of (\ref{quadx}) are used for computation.  It turns out that $x_1$ and $x_2$ are not canonically conjugate to each other. Using the Hamilton's equations for $x_1$ and $x_2$, 
\be\label{hamilton2}
\dot x_1 = \mu^\pm \, \f{\partial}{\partial x_2} {\cal H}^{\pm} ~, ~~~~~ \dot x_2 = - \mu^\pm \, \f{\partial}{\partial x_1} {\cal H}^{\pm} 
\ee
and the modified Raychaudhuri equation (\ref{modrai}), we obtain
\be
\mu^\pm = \pm 4 \pi G \, \sqrt{1 - 4 \alpha^2 x_2^{2}} ~.
\ee
%Thus, $x_2$ is not the conjugate momnentum of $x_1$. 
The conjugate momentum variable to $x_1$ can  be found using 
\be
p^\pm = \int  \f{\d x_2}{\mu^\pm} ~.
\ee

Let us first consider the negative root. The conjugate momentum turns out to be  
\be
p^- = \int \f{\d x_2}{\mu^-} = - \beta^{-1} \sin^{-1}(2 \alpha x_2) 
\ee
where we have defined $\beta = 8 \pi G \alpha$. Thus, $x_2$ for this root is 
\be
x_2 = - \f{\sin(\beta p^-)}{2 \alpha} ~.
\ee
Note that $x_2 = H$ is bounded between zero and $\pm 1/2\alpha$, with zero corresponding to the regime when $\rho$ vanishes. In the range of its principal values, we find  %The momentum  $p^-$ is a bounded function on the phase space:
\be
- \f{\pi}{2 \beta} ~ \leq ~ p^- ~ \leq 0 ~~~~ \mathrm{for} ~~~~ H \geq 0
\ee
and
\be
0 ~ \leq p^- \leq \f{\pi}{2 \beta}  ~~~~ \mathrm{for} ~~~~ H \leq 0 ~.
\ee
Using $p^-$, we can now rewrite the gravitational part of the Hamiltonian in (\ref{quadH-}) as:
\be \label{hamg1}
{\cal H}_g^- = - \f{3 V}{16 \pi G \alpha^2} \left(1 - \cos(\beta p^-)\right)  ~.
\ee
It is to be noted that the above gravitational 
Hamiltonian, due to the allowed range of $p^-$, is only valid for $2 \rcr \geq \rho \geq 0$. \\%$\rcr/2 \leq \rho \leq 0$. \\

Repeating this calculation for the positive root $\xi^+$, we obtain
\be
p^+ = \int \f{\d x_2}{\mu^+} =  \beta^{-1} \sin^{-1}(2 \alpha x_2) 
\ee
which yields
\be
x_2 =  \f{\sin(\beta p^+)}{2 \alpha} ~.
\ee
As in the case of the $x_2$ for the Hamiltonian with the negative root, $x_2$ lies in the range: $0 \leq |x_2| \leq 1/2\alpha$. However, 
unlike the previous case, $x_2$ does not vanish at $\rho = 0$, but at $\rho = 4 \rcr$. The range of $p^+$ is:
\be
0 ~ \leq p^+ \leq \f{\pi}{2 \beta}  ~~~~ \mathrm{for} ~~~~ H \geq 0 
\ee
and
\be
- \f{\pi}{2 \beta} ~ \leq ~ p^+ ~ \leq 0 ~~~~ \mathrm{for} ~~~~ H \leq 0
\ee
The gravitational part of the Hamiltonian in (\ref{H+}) turns out to be
\be \label{hamg2}
{\cal H}_g^+ = - \f{3 V}{16 \pi G \alpha^2} \left(1 + \cos(\beta p^+)\right)  ~.
\ee

Having obtained (\ref{hamg1}) and (\ref{hamg2}), we notice that the angles $\beta p^-$ and $\beta p^+$ do not belong to the same range of principal values for any given sign of Hubble rate. For $H \geq 0$, the maxima of the Hubble rate using  ${\cal H}_g^-$ is reached at $\beta p^- = - \pi/2$, and 
using ${\cal H}_g^+$ it is reached at $\beta p^+ =  \pi/2$. Similarly, for $H \leq 0$, the Hubble rate for the negative root reaches its maximum at  $\beta p^- =  \pi/2$, and for the positive root  it is reached at $\beta p^+ =  - \pi/2$. Further, the zero of $\beta p^-$ corresponds to the classical regime where the energy density vanishes, and that of $\beta p^+$ corresponds to the regime where the effects of repulsive effects of modified gravity lead to the bounce of the scale factor. It is convenient to introduce a new variable $p$, with a range
$- \pi/\beta \leq p \leq \pi/\beta$ and defined such that for $H \geq 0$,
\be
p := p^- ~~~~ \mathrm{for} ~~~~  ~ -\f{\pi}{2 \beta} \leq p \leq 0 ~,
\ee
and
\be
p := p^+ - \frac{\pi}{\beta}  ~~~~ \mathrm{for} ~~~~  ~ -\f{\pi}{\beta} \leq p \leq -\f{\pi}{2\beta} ~.
\ee
Similarly for $H \leq 0$,
\be
p := p^- ~~~~ \mathrm{for} ~~~~  ~ 0 \leq p \leq \f{\pi}{2 \beta}  ~,
\ee
and
\be
p := p^+ + \frac{\pi}{\beta}  ~~~~ \mathrm{for} ~~~~  ~ \f{\pi}{2\beta} \leq p \leq \f{\pi}{\beta} ~.
\ee
After expressing ${\cal H}_g^-$ and ${\cal H}_g^+$ in terms of $\beta p$, we can write the Hamiltonian  
for the entire range of $p$ as:
\be \label{Hcos}
{\cal H} = - \f{3 V}{16 \pi G\alpha^2} \left(1 - \cos(\beta p)\right) + {\cal H}_m \approx 0 ~.
\ee
This is the desired Hamiltonian in the form ${\cal H}_g$ + ${\cal H}_m$ in terms of the canonical variables $V$ and $p$. Incidentally, the 
resulting Hamiltonian can be identified with the one in the effective spacetime description of loop quantum cosmology  which is generally written in terms of 
the sine function \cite{aps3,vt},
%which using a trigometric identity takes the form
\be \label{Hsin}
{\cal H} = - \f{3 V}{16 \pi G \alpha^2} \sin^2(\beta p/2) + {\cal H}_m \approx 0 ~,
\ee
where we recall that $\alpha = (3/(32 \pi G \rho_c))^{1/2}$ and $\beta = 8 \pi G \alpha$. 
%which is obtained from (\ref{Hcos}) using a trigonometric identity. 
In loop quantum cosmology, $\beta$ encodes the area gap in the underlying quantum geometry which fixes the value of $\rcr$ in the modified Raychaudhuri 
equation. The effective Hamiltonian in loop quantum cosmology emerges for suitable semi-classical states after the polymer quantization of gravitational phase space variables has been performed 
\cite{vt}. In loop quantum cosmology, polymer 
quantization, which is inequivalent to Fock quantization, is tied to the usage of holonomies of connection as elementary variables for quantization. These holonomies result in 
trigonometric terms of momentum (which is related to connection) in the effective Hamiltonian \cite{as}.   
However, to obtain 
the Hamiltonian in eq.(\ref{Hsin}), we did not start from any input from loop quantum cosmology. We only assumed a modified Raychaudhuri 
equation (\ref{modrai}) which gives repulsive gravity with a $\rho^2$ correction with a positive sign and sought a canonical Hamiltonian 
linear in the matter Hamiltonian. Recall that in the classical case, the conjugate momentum to volume is proportional to the Hubble rate. Hence, the conjugate momentum in the present modified gravity scenario turns out to be an inverse trigonometric function of the classical momentum to the generalized coordinate $V$.  Polymerization thus emerges naturally. 

The Hamiltonian (\ref{Hcos}) is not 
restricted to the case of dust as a matter content but as in the case  of classical cosmology, is valid for matter Hamiltonian corresponding 
to any other equation of state. As in the classical case, it is straightforward to repeat the analysis and find that the same Hamiltonian is reached if one considers 
matter with different fixed equation of state. Finally, the modified Friedmann equation can be obtained by the vanishing of Hamiltonian ${\cal H}$, which is given by
\be
H^2 = \frac{8 \pi G}{3} \rho \left(1 - \frac{\rho}{4 \rcr}\right) ~.
\ee
This is the same Friedmann equation as in loop quantum cosmology for the spatially flat FRW model, with $4 \rcr$ corresponding to the bounce density 
\cite{ps06,aps3}. The dynamics resulting from the Hamiltonian has been extensively studied in loop quantum cosmology  \cite{as}. It turns out that 
the effective spacetime is geodesically complete and all strong curvature singularities are resolved in this modified gravity scenario \cite{ps09}.

\section{The Raychaudhuri equation with a quadratic attractive modification in energy density} 
We now consider a $\rho^2$ modification to the classical Raychaudhuri equation with a negative sign. In contrast to the modified 
Raychaudhuri equation considered in Sec. III which leads to a repulsive gravity irrespective of the equation of state once $\rho > \rcr$, 
the modification considered in this section makes gravity more attractive than in the classical GR. For the case of dust as matter, our starting point is 
the following modified Raychaudhuri equation: 
\be \label{modrai2}
\ddot a = - \f{4 \pi G}{3} \rho \left(1 + \f{\rho}{\rcr} \right) \, a ~.
\ee
As in the quadratic repulsive case, the modification to the Raychaudhuri equation is supposed to arise from a modified theory of gravity without changing  the number of 
degrees of freedom. The energy density scale $\rcr$ is to be determined from the underlying theory.

To determine the constant of the motion, as before we start with considering $x_1 = a$ and $x_2 = \dot a$. In terms of these variables, the 
modified Raychaudhuri equation (\ref{modrai2}) can be written as two first order equations:
\be
\dot x_1 = x_2, ~~~ \dot x_2 = -\f{4 \pi G}{3} \rho \left(1 +  \f{\rho}{\rcr}\right) x_1 ~.
\ee
Comparing the above set with eqs.(\ref{x1x2hojman}), we identify:
\be
A(x_1) = -\f{4 \pi G}{3} \rho \left(1 + \f{\rho}{\rcr}\right) x_1, ~~~~ \mathrm{and} ~~~~ B(x_2) = 1 ~.
\ee
Using eq.(\ref{Constant}) and the relation between energy density and matter Hamiltonian, we obtain a constant of the motion:
\be \label{hojmanH}
C(x_1,x_2) = \f{x_2^2}{2} - \f{4 \pi G}{3} \f{{\cal H}_m}{a^3} \left(1 + \f{1}{4 \rcr} \f{{\cal H}_m}{a^3} \right) x_1^2 ~.
\ee
It is straightforward to verify that, as in the classical and quadratic repulsive modification, $C(x_1,x_2)$ serves as a Hamiltonian. However, it is not in the form ${\cal H}_g + {\cal H}_m$. To obtain a Hamiltonian in the form linear in gravitational and matter Hamiltonian, we find the roots of the quadratic equation obtained from $C(x_1,x_2) \approx 0$:
\be\label{quadx2}
\xi^2 + \xi - \f{3}{32 \pi G} \f{H^2}{\rcr} = 0 ~, ~~~\mathrm{with} ~~~ \xi := \frac{\rho}{4 \rcr}
\ee
The resulting roots are:
\be
\xi_+ = \f{1}{2} (-1 + \sqrt{1 + 4 \alpha^2 H^2}), ~~~~ \mathrm{and} ~~~~ \xi_- = \f{1}{2} (-1 - \sqrt{1 + 4 \alpha^2 H^2}), ~~~ 
\mathrm{with} ~~~~ \alpha^2 := \f{3}{32 \pi G \rcr} ~.
\ee
The negative root implies $\rho < 0$. Hence, this root results in violation of weak energy condition and is not considered in the following discussion.\footnote{If we repeat the following analysis for this root,  instead of sine hyperbolic term, one obtains a cosine hyperbolic term in eq.(\ref{Hquadrep}) with $p = \beta^{-1} \sinh^{-1} (2 \alpha H)$.}
Unlike the case of quadratic repulsive modification to the Raychaudhuri equation, it is easily seen that the Hubble rate is not bounded in the present case. Another contrasting feature is that a single root $\xi_+$ spans the whole range of positive energy density. Thus, we expect the canonical Hamiltonian structure to be different than in the quadratic repulsive case. 

The Hamiltonian constraint $C(x_1,x_2) \approx 0$ results in equation of the form (\ref{quadrepeq}), with $\xi_\pm$ as roots given 
above. For the positive root which is allowed by weak energy condition, we obtain
\be\label{H-}
%{\cal H}^+ := - 
{\cal H} = \f{3 V}{8 \pi G \alpha^2}\,  \f{1}{2} (1 - \sqrt{1 + 4 \alpha^2 H^2}) + {\cal H}_m \approx 0 ~.
\ee
We thus obtain a Hamiltonian for the modified Raychaudhuri equation (\ref{modrai2}) in the form ${\cal H}_g + {\cal H}_m$. However, the gravitational part of the Hamiltonian is yet to be expressed in terms of the canonical variables.
%Let us choose $x_1 = V$ and $x_2 = H$ as the gravitational phase space variables satisfying $\{x_1, x_2\} = \mu$. 
As before, given the form of the Hamiltonian it is convenient to now choose $x_1 = V$ and $x_2 = H$. 
Computing Hamilton's equations and comparing with the modified Raychaudhuri equation (\ref{modrai2}), we find %that the Poisson bracket between $x_1$ and $x_2$ 
\be
\{V, H\} = \mu = - 4 \pi G \, \sqrt{1 + 4 \alpha^2 H^2} ~.
\ee
Thus, the conjugate momentum to $V$ turns out to be 
\be
p = \int \frac{\d x_2}{\mu} = - \beta^{-1} \sinh^{-1}(2 \alpha x_2) ~.
\ee
The Hubble rate diverges as the conjugate momentum $p$ diverges. This behavior is very similar to the case of the classical theory discussed in Sec. II, and is an indication of problem of singularities in this modified theory of gravity.

Expressing the Hamiltonian (\ref{H-}) in terms of the conjugate variables, we obtain 
\be\label{Hquadrep}
{\cal H} \, = \, - \frac{3 V}{8 \pi G \alpha^2} \, \sinh^2(\beta p/2) + {\cal H}_m \, \approx \, 0 ~, 
\ee
with $\alpha = (3/(32 \pi G \rho_c))^{1/2}$ and $\beta = 8 \pi G \alpha$. The above Hamiltonian results in the modified Friedmann equation:
\be
H^2 = \frac{8 \pi G}{3} \rho \left(1 + \frac{\rho}{4 \rcr}\right) ~.
\ee
We can now contrast the canonical Hamiltonian (\ref{Hquadrep}) with the Hamiltonian for the quadratic repulsive case (\ref{Hsin}). Instead of a trigonometric function of the classical conjugate momentum to volume, the Hamiltonian in quadratic attractive modification to the classical 
Raychaudhuri equation consists of a hyperbolic function. Thus, there is no polymerization. Physical solutions of this Hamiltonian are strikingly different from the one for the quadratic repulsive modification where the Hubble rate turned out to be universally bounded. In this modified gravity scenario, the dynamical solutions are singular. The resulting phase space structure and dynamics bears closer resemblance with the one in the classical Hamiltonian cosmology. 
Let us see whether the above modified gravity gravity scenario bears some resemblance with known models. 
It is interesting to note that a modified Friedmann equation with a similar correction arises on the four dimensional FRW brane in brane world scenarios if one ignores the contribution from the 5 dimensional bulk black hole, and the four dimensional cosmological constant vanishes \cite{roy}. In this case, $\rho_c$ gets identified with the brane tension. The covariant Raychaudhuri equation on the brane then agrees with modified Raychaudhuri equation (\ref{modrai2}).  %Therefore, the modified gravity Hamiltonian obtained above does not capture the brane world dynamics. 
Incidentally,  the Hamiltonian constraint (\ref{Hquadrep}) can also be considered as originating from Euclideanized version of (\ref{Hsin}) or equivalently loop quantum cosmology Hamiltonian constraint using an imaginary value of the conjugate momentum $p$ (or a complex Ashtekar connection).

\section{The Raychaudhuri equation with a cubic repulsive modification in energy density}
Let us consider the case of a repulsive modification to the classical Raychaudhuri equation with a $\rho^3$ correction. As before, the 
modification is assumed to not introduce any additional degrees of freedom. For dust as matter, the 
classical Raychaudhuri equation modifies as follows:%as follows without changing the degrees of freedom:
\be\label{modraycubic1}
\ddot a = - \f{4 \pi G}{3} \rho \left(1 - \frac{\rho^2}{\rho_c^2}\right) a ~.
\ee
Using $x_1 = a$ and $x_2 = \dot a$, this equation can be written as two first order differential equations,
\be
\dot x_1 = x_2, ~~~~ \dot x_2 = -\frac{4 \pi G}{3} \rho \left(1 - \frac{\rho^2}{\rho_c^2}\right) x_1 ~.
\ee
Comparing with eq.(\ref{x1x2hojman}), we can identify
\be
A(x_1) = -\frac{4 \pi G}{3} \rho \left(1 - \frac{\rho^2}{\rho_c^2}\right) x_1, ~~~ \mathrm{and} ~~~ B(x_2) = 1 ~.
\ee
The constant of the motion using (\ref{Constant}) turns out to be
\be
C(x_1,x_2)  = \frac{x_2^2}{2} - \frac{4 \pi G}{3} \rho \left(1 - \frac{\rho^2}{7 \rho_c^2}\right) x_1^2 ~.
\ee
As in the case of $\rho^2$ modifications to the Raychaudhuri equation, $C(x_1,x_2)$ serves as a Hamiltonian which vanishes due to general covariance of the theory. To obtain a Hamiltonian in the form 
${\cal H}_g + {\cal H}_m$, we need to solve a cubic equation obtained from $C(x_1,x_2) \approx 0$. This equation is 
% The Hamiltonian consraint takes the form 
\be \label{cubic}
\zeta^3 - \zeta + \sigma^2 H^2 = 0 ~.
\ee
Here we have defined $\zeta = \rho/\sqrt{7} \rho_c$, and $\sigma^2 = 3/(8 \sqrt{7} \pi G \rho_c)$. From this equation we see that the 
maxima of $H^2$ occurs at $\zeta^2 = 1/3$. For $\rho \geq 0$, the maximum value 
of $|H|$ is obtained at $\zeta = 1/\sqrt{3}$, and this value is 
\be
H^2_{\rm{max}} = \f{2}{3 \sqrt{3} \sigma^2} ~.
\ee  
Thus, as in the quadratic repulsive modifications studied in Sec. III, the cubic repulsive modification results in a bounded Hubble rate. 
Insights from studies in loop quantum cosmology where the Hubble rate is also universally bounded \cite{cs2,ps09}, whose Hamiltonian can be identified with the 
quadratic repulsive case, suggest that dynamics is singularity free \cite{ps16}. The Hubble rate vanishes in the high curvature regime 
at $\rho = \sqrt{7} \rho_c$ where the scale factor of the universe bounces avoiding the big bang singularity.

Eq. (\ref{cubic}) has three real roots, which are:
\be \label{zeta1}
\zeta_1 = \f{2}{\sqrt{3}} \cos\left(\f{1}{3} \cos^{-1}\left(-\f{3 \sqrt{3} \sigma^2 H^2}{2}\right)\right) ~,
\ee
\be
\zeta_2 = \f{2}{\sqrt{3}} \cos\left(\f{1}{3} \cos^{-1}\left(-\f{3 \sqrt{3} \sigma^2 H^2}{2}\right) + \f{2 \pi}{3} \right) ~,
\ee
and 
\be \label{zeta3}
\zeta_3 = \f{2}{\sqrt{3}} \cos\left(\f{1}{3} \cos^{-1}\left(-\f{3 \sqrt{3} \sigma^2 H^2}{2}\right) + \f{4 \pi}{3}\right) ~.
\ee
These roots allow us to write a Hamiltonian constraint starting from $C \approx 0$:
\be \label{cubicrepeq}
{\cal H} = - \sqrt{7} \zeta_i \,\rcr V + {\cal H}_m \approx 0, ~~~ (i = 1,2,3), 
\ee
which is analog of eq.(\ref{classicalHamrel}) in the classical case and eq.(\ref{quadrepeq}) in the quadratic modifications to 
Raychaudhuri equation.

It is instructive to take the limit $\alpha H \rightarrow 0$ to gain some insights on these roots. In this limit, the roots can be expanded 
as 
\be \label{zeta1approx}
\zeta_1 \simeq 1 - \f{\sigma^2 H^2}{2} - \f{3 \sigma^4 H^4}{8} - \f{\sigma^6 H^6}{2} - \f{105 \sigma^8 H^8}{128} + O(\sigma^{10} H^{10}) ~,
\ee
\be
\zeta_2 \simeq - 1 - \f{\sigma^2 H^2}{2} + \f{3 \sigma^4 H^4}{8} - \f{\sigma^6 H^6}{2} +  \f{105 \sigma^8 H^8}{128} + O(\sigma^{10} H^{10}) ~,
\ee
and 
\be \label{zeta3approx}
\zeta_3 \simeq \sigma^2 H^2 + \sigma^6 H^6 + 3 \sigma^{10} H^{10} + O(\sigma^{13} H^{13}) ~.
\ee
Note that in the absence of $\rho^3/\rcr^3$ modification,  i.e. in the classical theory $\zeta = \sigma^2 H^2$. Thus, the root $\zeta_3$ 
captures the classical limit in the regime where Hubble rate is vanishingly small. 
The root $\zeta_1$ captures the bounce regime where the Hubble rate vanishes. On the other hand, the root $\zeta_2$ is always negative, and 
results in violation of weak energy condition. In the following, we will only consider roots $\zeta_1$ and $\zeta_3$ as 
it is only for these the energy density is positive. The $\zeta_1$ root provides dynamics in the range of $\rho = \sqrt{7} \rho_c$, the 
maximum value of energy density, till $\rho = \sqrt{7} \rho_c/\sqrt{3}$ where Hubble rate attains its maximum value. The $\zeta_3$ root 
covers dynamics in the lower range of energy density, from $\rho = \sqrt{7} \rho_c/\sqrt{3}$ till vanishing energy density.

We first consider the Hamiltonian corresponding to the $\zeta_1$ root (eq.(\ref{zeta1})). The vanishing of $C(x_1,x_2)$ yields
\be \label{Hzeta1}
{\cal H}(x_1,x_2) = - \f{3 x_1 \zeta_1}{8 \pi G \sigma^2} +  {\cal H}_m \approx 0
\ee
where we have chosen $x_1 = V$ and $x_2 = H$, satisfying $\{x_1,x_2\} = \mu_{\zeta_1}$. The Hamilton's equation for $x_1$ is:
\be
\dot x_1 = \mu_{\zeta_1} \f{\partial {\cal H}}{\partial x_2} = \frac{3 x_1 x_2}{4 \pi G} \mu_{\zeta_1} \f{\sin\left(\f{1}{3} \cos^{-1} 
\left(-\chi^2\right)\right)}{\sqrt{1 - \chi^4}} 
\ee 
where $\chi^2$ is defined as $\chi^2 =  \f{3 \sqrt{3} \sigma^2 H^2}{2}$. % and $\mu$ satisfies $\{x_1,x_2\} = \mu(x_1,x_2)$. 
Using $\dot V = 3 V H$, we find that 
\be
\mu_{\zeta_1} = 4 \pi G \f{\sqrt{1 - \chi^4}}{\sin\left(\f{1}{3} \cos^{-1} \left(-\chi^2\right)\right)} ~.
\ee
Thus, $x_1$ and $x_2$ are not conjugate variables.  The conjugate momentum to $x_1$ can be found by using 
\be\label{pintegral}
p_{\zeta_1} = \nonumber  \int \, \f{\d H}{\mu_{\zeta_1}} ~.
\ee
Note that $\mu_{\zeta_1}$ is a bounded function, with its 
maximum value equal to $8 \pi G$ attained when $\sigma H \rightarrow 0$, and the minimum value equal to zero is reached when $\sigma |H| = 
\sqrt{2/3\sqrt{3}}$. The integral yields an intricate relation between the conjugate momentum and the Hubble rate:
\ba\label{pxiexact2}
p_{\zeta_1} &=& \nonumber \frac{1}{4 \pi G} \frac{3^{1/4}}{5 \sqrt{2} \sigma}  \Bigg[\chi^{-1} \sqrt{1 + e^{2 i 
\cos^{-1}(-\chi^2)}} e^{-\frac{i}{3} 
\cos^{-1}(-\chi^2)} 
 \bigg[-e^{\frac{2 i}{3} \cos^{-1}(-\chi^2)}\, _2F_1\left(\frac{5}{12},\frac{1}{2};\frac{17}{12};e^{-2 i \cos^{-1}(-\chi^2)}\right) \\
&& ~~~~~~~~~~~~~~~~~~~~~~~~~~~~~~~~~~~~~~~~~~~~~~~~~~~~~~ + ~~~ 5 \, _2F_1\left(\frac{1}{12},\frac{1}{2};\frac{13}{12};e^{-2 i 
\cos^{-1}(-\chi^2)}\right) \bigg]\Bigg] ,
\ea
where let us recall that $\sigma = (3/(8 \sqrt{7} \pi G \rho_c))^{1/2}$.  
The conjugate momentum is a hypergeometric function of the inverse trigonometric function of the classical conjugate momentum $H$. 
Taking the allowed limits of the Hubble rate, it can be checked that it is real and finite.

\vskip0.5cm
We now consider the root $\zeta_3$ (eq.(\ref{zeta3})). 
In this case, the Hamiltonian constraint in terms of $x_1 = V$ and $x_2 = H$ is 
\be \label{Hzeta3}
{\cal H}(x_1,x_2) = - \f{3 x_1 \zeta_3}{8 \pi G \sigma^2} +  {\cal H}_m \approx 0 ~.
\ee
Using Hamilton's equations, we find that $\mu_{\zeta_3}$ is given by 
%a similar calculation for $\mu$ yields
\be
\mu_{\zeta_3} = 4 \pi G \f{\sqrt{1 - \chi^4}}{\cos\left(\frac{\pi}{6} - \f{1}{3} \cos^{-1} \left(-\chi^2\right)\right)} ~
\ee
where $\chi^2 = \f{3 \sqrt{3} \sigma^2 H^2}{2}$. We find that $\mu_{\zeta_3}$  
 is a bounded function with a maximum and minimum values obtained at $\sigma H  = 0$ and $\sigma |H| = 
\sqrt{2/3\sqrt{3}}$ respectively.
%equal to zero at $\sigma H = {\sqrt{2/3\sqrt{3}}}$, and the minimum value equal to 
% $-1/2$ at $\alpha H = 0$. 
 The conjugate momentum to $x_1$ turns out to be 
\ba
p_{\zeta_3} &=& \nonumber -\frac{1}{4 \pi G} \frac{3^{1/4}}{5 \sqrt{2} \sigma}  \Bigg[(-1)^{1/3} \chi^{-1} \sqrt{1 + e^{2 i 
\cos^{-1}(-\chi^2)}} e^{-\frac{i}{3} 
\cos^{-1}(-\chi^2)} \\&& \nonumber
 \bigg[e^{\frac{2 i}{3} \cos^{-1}(-\chi^2)}\, _2F_1\left(\frac{5}{12},\frac{1}{2};\frac{17}{12};e^{-2 i \cos^{-1}(-\chi^2)}\right) 
%&& ~~~~~~~~~~~~~~~~~~~~~~~~~~~~~~~~~~~~~~~~~~~~~~~~~~~~~~ 
+ ~ 5 (-1)^{1/3}\, _2F_1\left(\frac{1}{12},\frac{1}{2};\frac{13}{12};e^{-2 i 
\cos^{-1}(-\chi^2)}\right) \bigg]\Bigg] \\
\ea
which like $p_{\zeta_1}$ is  real and finite in the allowed range of Hubble rate.

The conjugate momenta $p_{\zeta_1}$ and $p_{\zeta_3}$ capture the dynamics in the phase space for $\sqrt{7} \rho_c \geq \rho \geq \sqrt{7} 
\rho_c/\sqrt{3}$ and $\sqrt{7} \rho_c/\sqrt{3} \geq \rho \geq 0$ respectively. Unfortunately, 
their complicated form forbids us to express the roots $\zeta_1$ (eq.(\ref{zeta1})) and $\zeta_3$ (eq.(\ref{zeta3})) in terms of 
$p_{\zeta_1}$ 
and $p_{\zeta_3}$. Nevertheless, valuable information is gained by plotting $p_{\zeta_1}$ and $p_{\zeta_3}$ versus the roots $\zeta_1$ and $\zeta_3$, and the Hubble rate. These plots are shown in Fig. 1, which depict  periodic relationships between the conjugate momenta and the corresponding roots, and the Hubble rate. Their periodic variations turn out to be very similar to the one for the quadratic repulsive case studied in Sec. III. To understand this in detail and to compare with the quadratic repulsive case, let us denote the conjugate momenta $p_{\zeta_1}$ and $p_{\zeta_3}$ as $p$, where it is understood that contributions to $p$ come from the former momenta in their respective ranges. It turns out that the variations are periodic in $p$. 
%where $\eta = 4 \pi G \sigma$ and $\sigma = (3/(8\sqrt{7} \pi G \rho_c))^{1/2}$. 
The period of an oscillation for the entire allowed range of the Hubble rate is approximately $p = 2.62/\eta$  where $\eta = 4 \pi G \sigma$ and $\sigma = (3/(8\sqrt{7} \pi G \rho_c))^{1/2}$. Note that in the quadratic repulsive case, this period is $p = 2 \pi/\beta$, with $\beta = 8 \pi G \alpha$ and $\alpha = (3/(32 \pi G \rho_c))^{1/2}$. The maximum value of $|\sigma H|$ in the cubic repulsive case is reached at approximately $p = (\pi/6 \eta)$. In contrast, the maximum value of $|\alpha H|$ in the quadratic repulsive case is reached at $p = (\pi/2 \beta)$. 

Let us see the way the criticial energy density in the quadratic and cubic repulsive cases is related to the period of oscillation. In the quadratic repulsive case, the value of the conjugate momentum $p$ where $\alpha H$ becomes maximum lies exactly in the middle of the half-period of oscillation. At this value of $p$, which is $p = \pi/2 \beta$ the energy density reaches $2 \rho_c$. The maximum value of energy density, $\rho = 4 \rho_c$ is reached at $p = \pm \pi/\beta$. %Thus, energy density goes from zero to its maximum value in the period $\pi/\beta$ for the quadratic repulsive case. 
The relationship between the period $P$ and the $\rho_c$ in the quadratic repulsive case turns out to be 
\be
%P = \left(\frac{2 \pi}{3 G}\right)^{1/2} \rho_c^{1/2} ~ \approx 1.45 \left(\frac{\rho_c}{G}\right)^{1/2}  ~,
P = \frac{2 \pi}{\beta} \approx  1.45 \left(\frac{\rho_c}{G}\right)^{1/2} .
\ee

In constrast, in the cubic repulsive case the value of $p$ at which $|\sigma H|$ attains its maximum does not lie exactly in the middle of the half-period, but at a little less value. At this value which is approximately $p = \pi/(6 \eta)$, energy density becomes $\rho = \sqrt{7/3} \rho_c$. The maximum value of  energy density, $\rho = \sqrt{7} \rho_c$, is reached at $p = \pm 1.31/\eta$. This is evident in the left plot of Fig. 1. In the cubic repulsive case, the period of oscillation $P$ is related to critical density $\rho_c$ as 
\be
P = \frac{2.62}{\eta} \approx 0.98 \left(\frac{\rho_c}{G}\right)^{1/2} .
\ee
Hence the period of oscillation in the quadratic repulsive case is approximately 1.5 times that in the cubic repulsive case.

%In the cubic repulsive case, the 
%corresponding value of the energy density in terms of the critical energy density $\rho_c$ when $|\sigma H|$ attains its maximum does not lie exactly in the middle of the half-period in the cubic repulsive modification, but at a little less value. In the quadratic repulsive case, this value lies in the middle of the half-period. 
In Fig. 1, we also see the periodic behavior of the roots $\zeta_i$ with respect to the conjugate momenta  for two cycles ranging in all the allowed values of the Hubble rate. Again, the behavior is quite similar to that of the quadratic repulsive case, except with a shorter period which we calculated above. In the quadratic repulsive case, two such cycles cover $p = 4 \pi/\beta$.  Thus we find, quadratic repulsive and cubic repulsive modifications result in a qualitatively similar, but  quantitatively distinct periodic behaviors of the corresponding roots in terms of conjugate momentum. Another way to look at these similarities is via the expansion of $\zeta$'s around the bounce point and the classical regimes 
considered in eqs.(\ref{zeta1approx}) and (\ref{zeta3approx}). If one considers the expansion of corresponding roots ($\xi$'s) in the quadratic repulsive case, then one finds that they have very similar expansions in terms of the Hubble rates in the bounce and the classical regimes.

\begin{figure}[tbh!]
\includegraphics[width=7cm]{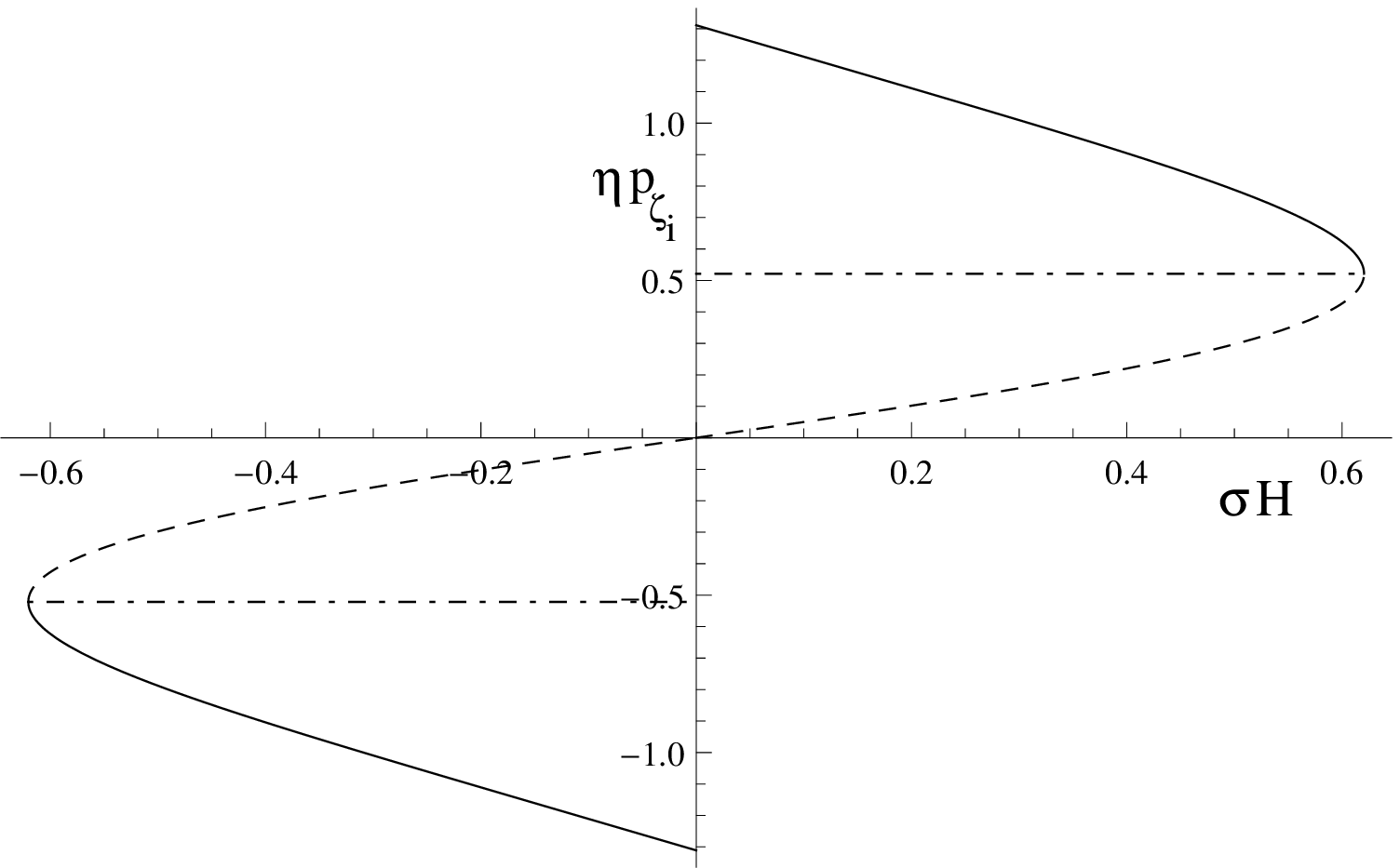}
\hskip0.5cm
\includegraphics[width=7cm]{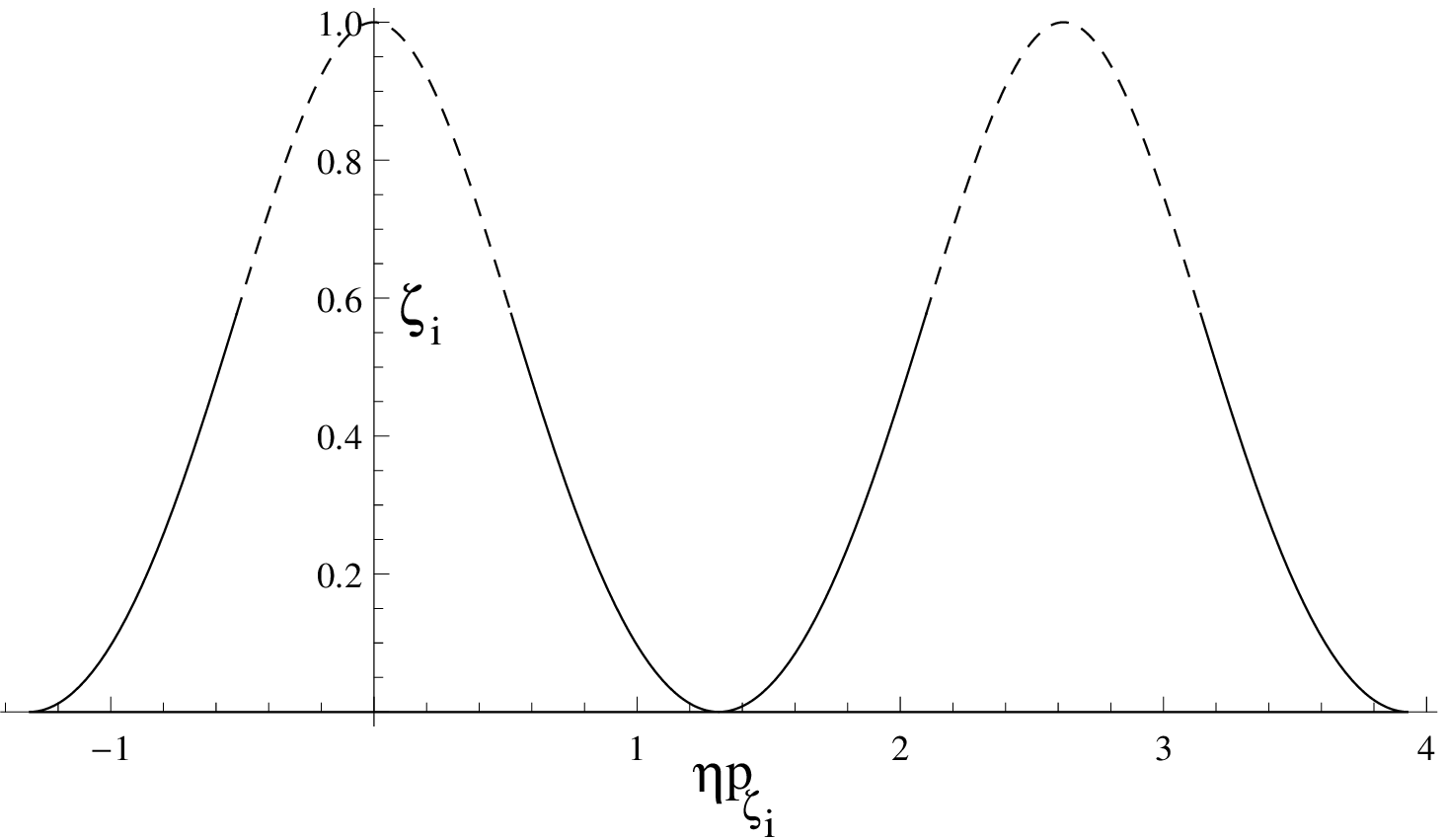}
\caption{The periodic relationship of $\eta p_{\zeta_i}$ (where $i = 1,3$) is shown versus the corresponding roots $\zeta_i$ and the Hubble rate. Here $\eta = 4 \pi G \sigma$ and $\sigma = (3/(8\sqrt{7} \pi G \rho_c))^{1/2}$. The solid curve indicates $p_{\zeta_1}$ (and $\zeta_1$) and the dashed curve corresponds to $p_{\zeta_3}$ (and $\zeta_3$). In the left plot, the dot-dashed curve shows the value of $\eta p_{\zeta_i}$ at which $\rho = \sqrt{7/3} \rho_c$ and the magnitude of $\sigma H$ attains its maximum value. The right plot shows two cycles for the entire allowed range of the Hubble rates. In each cycle, the total period is approximately $\eta p_{\zeta_i} = 2.62$.} %In the right plot, the period of variation of $\sigma H$ is approximately 1.31 in $\eta p_{\zeta_i}$ for positive or negative values of the Hubble rate.}
\end{figure}

The Hamiltonians (\ref{Hzeta1}) and (\ref{Hzeta3}) which capture the evolution in different ranges of energy density 
give us the desired Hamiltonian structure with Hamiltonian as linear combination of gravitational and matter parts. As in the earlier cases, we note that though we 
started with dust as matter, the final Hamiltonian is valid for a general matter Hamiltonian. 
 We notice that as in the case of 
repulsive quadratic modification, $\rho^3$ modification to the classical Raychaudhuri equation yields 
a canonical phase space structure which is disjoint in conjugate momenta for the entire allowed range of energy density. The conjugate 
momenta again involves inverse trigonometric functions of the Hubble rate, albeit relatively complicated ones. These are  hypergeometric functions of the inverse trigonometric functions, which as discussed above and depicted in Fig. 1 share very similar periodic features with the quadratic repulsive case.  The periodic behavior of the roots $\zeta_{(1,3)}$ with respect to the conjugate momenta implies that as in the quadratic repulsive case, Hamiltonian is also periodic in the conjugate momentum. Due to the existence of periodicity of the Hamiltonian in the conjugate momentum in a more general form than the simple 
trigonometric function in the quadratic repulsive case, which is linked to polymerized phase space, one can view the resulting phase space as of `generalized polymerized' form.  

%presence of inverse 
%trigonometric functions as in the quadratic repulsive case which is linked to polymerized phase space, but now in a more general form, one can view the resulting phase space as of `generalized polymerized' form.  

Finally, the modified Friedmann equation emerging from the vanishing of the Hamiltonian in this modified gravity scenario is
\be
H^2 = \frac{8 \pi G}{3} \rho \left(1 - \frac{\rho^2}{7 \rcr^2}\right) ~.
\ee
Unlike the quadratic repulsive modification,  phenomenological implications resulting from this and the modified Raychaudhuri equation 
(\ref{modraycubic1}) of this new non-singular Hamiltonian cosmology are yet to be studied. We expect many features to be qualitatively 
similar to the quadratic repulsive case, however quantitative details and predictions will be different.

\section{The Raychaudhuri equation with a cubic attractive modification in energy density}
 The cubic repulsive modification to the classical Raychaudhuri equation causes gravity to be more attractive than any of the 
previous cases for modified gravity scenarios studied so far in this manuscript. The modified Raychaudhuri equation for dust as matter is assumed to be of the following form:
\be \label{cubicattray}
\ddot a = - \f{4 \pi G}{3} \rho \left(1 + \frac{\rho^2}{\rho_c^2}\right) a ~.
\ee
As in all other considered modifications in our analysis, general covariance is assumed to remain unchanged and no new degrees of freedom are added in this modified 
gravity scenario.

Starting with $x_1 = a$ and $x_2 = \dot a$, we can rewrite eq.(\ref{cubicattray}) as two first order equations:
\be
\dot x_1 = x_2, ~~~~ \dot x_2 = -\frac{4 \pi G}{3} \rho \left(1 + \frac{\rho^2}{\rho_c^2}\right) x_1 ~.
\ee
This set is equivalent to (\ref{x1x2hojman}) for 
\be
A(x_1) = -\frac{4 \pi G}{3} \rho \left(1 + \frac{\rho^2}{\rho_c^2}\right) x_1, ~~\mathrm{and} ~~~ B(x_2) = 1 ~.
\ee
Using $A(x_1)$ and $B(x_2)$ in eq.(\ref{Constant}) we find the constant of the motion in this modified gravity as 
\be
C(x_1,x_2)  = \frac{x_2^2}{2} - \frac{4 \pi G}{3} \rho \left(1 + \frac{\rho^2}{7 \rho_c^2}\right) x_1^2 ~.
\ee
This constant of the motion serves as a Hamiltonian yielding Hamilton's equations consistent with modified Raychaudhuri equation (\ref{cubicattray}) for $x_1$ and $x_2$ as conjugate variables. However, as for all other previous cases, it is not a linear combination of gravitational and matter Hamiltonians. Using our strategy of imposing $C \approx 0$, to find the desired Hamiltonian, we are led to the following cubic equation:
%The Hamiltonian consraint takes the form 
\be \label{cubic}
\zeta^3 + \zeta - \sigma^2 H^2 = 0 ~.
\ee
Here $\zeta$ and $\sigma$ are defined as in Sec. V, $\zeta := \rho/\sqrt{7} \rho_c$, and $\sigma^2 = 3/8 \sqrt{7} \pi G \rho_c$. The above equation permits only 
one real root which allows positive energy density. This single root $\zeta$ is 
\be
\zeta = \f{2}{\sqrt{3}} \sinh\left(\frac{1}{3} \sinh^{-1} \chi^2\right) ~.
\ee
The vanishing of $C$ yields the following Hamiltonian constraint in the form ${\cal H}_g + {\cal H}_m$,
It satisfies the Hamiltonian constraint obtained from $C \approx 0$ in the form 
\be
{\cal H} = - \f{3 x_1 \zeta}{8 \pi G \sigma^2} +  {\cal H}_m \approx 0
\ee
where $x_1 = V$ and $x_2 = H$. This equation is equivalent to (\ref{cubicrepeq}) for $\zeta$ identified with $\zeta_i$. To find the conjugate phase space variables, we analyze Hamilton's equations. The Hamilton's equation for $x_1$ is:
\be
\dot x_1 = \mu \f{\partial {\cal H}}{\partial x_2} = -\frac{3 x_1 x_2}{4 \pi G} \mu 
\f{\cosh\left(\f{1}{3} \sinh^{-1} \left(\chi^2\right)\right)}{\sqrt{1 + \chi^4}}
\ee 
where as before $\chi^2 = \f{3 \sqrt{3} \sigma^2 H^2}{2}$. Consistency of Hamiltonian evolution requires 
\be
\mu = -4 \pi G \f{\sqrt{1 + \chi^4}}{\cosh\left(\f{1}{3} \sinh^{-1} \left(\chi^2\right)\right)} ~.
\ee
The conjugate variable to $x_1$ can be found by the following integration
\be
p = \int \frac{\d H}{\mu} ~,
\ee
which yields 
\ba
p &=& \nonumber -\frac{1}{4 \pi G} \frac{3^{1/4}}{5 \sqrt{2} \sigma} \chi^{-1} e^{-\frac{1}{3} \sinh^{-1} \chi} \sqrt{1 - e^{2 \sinh^{-1} 
\chi}} \bigg[5 \, _2F_1\left(\frac{1}{12}, \frac{1}{2}; \frac{13}{12}; e^{2 \sinh^{-1} \chi}\right) \\
&& ~~~~~~~~~~~~~~~~~~~~~~~~~~~~~~~~~~~~~~~~~~~~~ + ~~~ e^{\frac{2}{3} \sinh^{-1} \chi}  \,
_2F_1\left(\frac{5}{12}, \frac{1}{2}; \frac{17}{12}; e^{2 \sinh^{-1} \chi}\right) \bigg] ~,
\ea
with $\sigma = (3/(8 \sqrt{7} \pi G \rho_c))^{1/2}$. 
Let us compare the resulting canonical phase space structure with the previous cases. As in quadratic attractive modification, the conjugate momentum involves 
inverse hyperbolic functions of Hubble rate and there is no polymerization. But as in the cubic repulsive case, one obtains hypergeometric functions which make expressing $\zeta$ in terms of $p$ complicated.  
In another similarity to the quadratic attractive case, there is only one root, $\zeta$, which covers the whole positive range of energy density. The canonical phase space structure is hence quite different from the repulsive modified gravity cases where there are two physical roots. It is straightforward to see that the Hubble rate in this case 
is unbounded, which is also transparent from the modified Friedmann equation following from the vanishing of Hamiltonian,
\be
H^2 = \frac{8 \pi G}{3} \rho \left(1 + \frac{\rho^2}{7 \rcr^2}\right) ~.
\ee
This equation, as is the Hamiltonian, is valid for matter with any equation of state. A divergence in energy density causes Hubble rate to diverge and the resulting spacetime is 
geodesically incomplete. The approach to singularity occurs faster than the classical and quadratic attractive cases.

\section{Discussion}

Let us begin with summarizing the main goal and steps of our procedure. 
We have provided a systematic method to determine the canonical Hamiltonian directly from  the equations of motion
in spatially flat isotropic and homogeneous cosmological models in modified  gravity without any information about the Lagrangian. We asked  
what type of 
canonical structures of the cosmological models in modified theories produce the total Hamiltonian as a linear 
combination of gravity and matter Hamiltonians, ${\cal H}_g + {\cal H}_m$, without violation of conservation law of matter-energy and 
general covariance.  We  considered 
modifications to the Raychaudhuri equation involving 
quadratic and cubic terms of energy density which make gravity repulsive above a curvature scale for matter satisfying strong energy 
condition, or more attractive than in GR. Modifications are assumed to be such that no new degrees of freedom are added. Using these conditions we obtain the Hamiltonian and canonical phase 
space structure for all the considered 
modified cosmological theories. In the quadratic repulsive modification, our approach can be viewed as a reverse procedure to obtain the 
effective canonical Hamiltonian  in loop quantum cosmology. For other modifications, canonical Hamiltonian structures found here are new and have been 
investigated for the first time in literature.

Our starting point is finding of a constant of the motion $C$. This constant of the motion serves as an intermediate Hamiltonian in our 
procedure. It does not have the distinction of being the desired Hamiltonian because it does not comply with our requirement that the Hamiltonian 
be a linear combination of gravitational and matter parts. At this point it becomes crucial to use the vanishing of the (intermediate) Hamiltonian ($C 
\approx 0$), a result of general covariance and phase space variables transforming as scalars under time reparameterization, to obtain the desired Hamiltonian and the canonical phase space structure. As a consequence of our three requirements -- the 
constancy of $C$, the vanishing of $C$ which gives physical space of solutions, and, ${\cal H} = {\cal H}_g + {\cal H}_m$, we find starting from Raychaudhuri equation and its modifications that 
${\cal H} = {\cal H}_m - \rho a^3$, where $\rho$ is a root of the equation $C \approx 0$. Once the Hamiltonian is obtained in this form, we derive the canonical phase space structure. This is the main summary of our procedure. 
Note that our approach is based on the relation between the Hamiltonian and the existence of constants of motion which 
can be $2N-1$ at most for $N$ degrees of freedom. But all problems always have $2N$ constants of motion, the $2 N$ initial conditions. This 
is taken care of by assuming that at most $2 N - 1$ constants of motion have to be time independent and isolating, i.e. allowing for 
reduction of the dimensionality of the phase space. 

What modifications to the Raychaudhuri equation tell us about the underlying canonical Hamiltonian structure? In the quadratic repulsive case, 
the modified phase space of gravity is a polymerized phase space which is characteristic of models in loop quantum gravity. The 
 canonical momentum turns out to be an inverse trigonometric 
function of Hubble rate. The canonical Hamiltonian can be identified with the effective Hamiltonian in loop quantum cosmology if one 
appropriately identifies $4 \rho_c$ with the bounce density in loop quantum cosmology.   The repulsive cubic modification results in a new nonsingular Hamiltonian cosmology 
whose gravitational phase space has a `generalized polymerized' structure with momenta as hypergeometric functions of inverse trigonometric 
functions of Hubble rate. Both of the repulsive modifications are found to 
yield a bounded Hubble rate. Here it is worth noting that investigations on resolution of strong curvature singularities in loop quantum cosmology \cite{ps09}, show that these are generically resolved. This conclusion immediately extends to the quadratic repulsive case, and, similar conclusions follow for the cubic repulsive case \cite{ps16}. In contrast, attractive modifications result in a non-polymerized gravitational phase 
space. In both of the cases considered here, Hubble rate is not bounded and singularities persist for matter which does not violate weak energy condition. The quadratic attractive modification results in modified gravity which has similarities with the brane world scenarios \cite{roy}.

Let us discuss some of the  directions in which our procedure can be extended and generalized. 
We have focused in this manuscript on the Hamiltonian of the modified gravity scenarios. The question of what is the corresponding covariant action for these canonical Hamiltonians remains to be addressed. It has been earlier shown that effective dynamical equation of loop quantum cosmology can be obtained from a covariant generalized Palatini action $f({\cal R})$, where $f({\cal R})$ is an infinite series in curvature scalar of the connection ${\cal R}$, with metric-connection compatibility not enforced \cite{action}.  It will be interesting to understand this result in light of the inverse procedure provided in this manuscript for the repulsive modified gravity cases. The action obtained in Ref. \cite{action}, can be considered as an action for the quadratic repulsive case in our analysis. However, one needs to understand the relationship between the canonical Hamiltonian structure and this action in more detail. Whether or not such an action exists for the cubic repulsive case is an open question. These 
exercises promise to  
give valuable insights on modified gravity scenarios where degrees of freedom do not change, as considered in our analysis, and investigations starting from higher order actions. Similarly, it will be interesting to 
generalize our procedure to include non-minimally coupled matter and modifications allowing change in the degrees of freedom. 

Understanding the uniqueness of the Hamiltonians obtained in this manuscript is an interesting avenue to explore. In particular, an important question is the following. Do independent and very different procedures of obtaining the Hamiltonian from the modified Raychaudhuri equations agree? The answer seems to be positive. An independent method from what is put forward in our manuscript has its roots in the study of non-linear dynamical systems (with variable coefficients) which has received attention for well over two centuries. In this context, Nucci and Leach 
have recently resurrected an old method of Jacobi to derive the Lagrangian description of second-order dynamical system, or even a system of such ordinary differential equations \cite{leach}. This direction has also been explored in earlier works by Rao \cite{rao}, and Whittaker \cite{whittaker}.  It is interesting to use this technique connecting the Jacobi's Last Multiplier method with the Lagrangian formulation of differential equations and 
determine the Lagrangian of the modified cosmological equations as considered in this manuscript. Using the standard Legendre transformation we deduce the corresponding Hamiltonian once the Lagrangian is found. Following this independent 
procedure going back to Jacobi for the modified Raychaudhuri equations studied in this manuscript, the Hamiltonians turn out to be of the same form as obtained in our analysis \cite{ss2}. In our ongoing work, this analysis is extended to variants of the modified cosmological models studied here, and the precise nature of inter-relation between the Hamiltonian deduced from our procedure and the Lagrangian function obtained using Jacobi's Last Multiplier method is established. Results from our ongoing analysis show that very distinct methods lead to same Hamiltonians for the modified gravity scenarios.

It should be noted that starting from the Raychaudhuri equation is not fundamental to find the Hamiltonian in our procedure. Instead, one 
can use Friedmann equation and its modifications to obtain canonical Hamiltonian structure. One can start with a constant of the motion, $C = {\cal H}_m - \rho a^3$,  which follows independently from conservation of the matter-energy. If one assumes 
that $\rho$ is a solution of the Friedmann equation of the classical or modified gravity scenario, $C$ has the distinction of being the desired Hamiltonian of 
the model satisfying our requirements of ${\cal H} \approx 0$ and ${\cal H} = {\cal H}_m + {\cal H}_g$. It only remains to identify the 
canonical gravitational phase space variables. This can be done following considerations given in Secs. II-VI for classical and modified 
gravity scenarios. This procedure will be used in other cosmological models in an upcoming work \cite{ss2}.

It can be interesting to compare our approach with the historic problem of finding the force law of gravity in Newtonian mechanics. Given a 
potential, one can always determine the trajectory it leads to. Motion under the Newtonian potential and the harmonic oscillator potential 
has many extraordinary properties. They are regular problems in which all bounded planar orbits are closed, whereas it is known that most 
potentials do not even lead to closed trajectories. In a converse sense, the mentioned atypical properties of the motion is the key for 
understanding the dynamical behavior of Hamiltonian systems. The Kepler problem and the isotropic harmonic oscillator have the only radial 
potentials in which all finite orbits are in plane and closed in agreement with the well known Bertrand's theorem. In a similar way, precision astronomical observations can provide valuable lessons to understand the Hamiltonian structure of the modified gravity scenarios describing our universe at large spacetime curvature. On this we note that ongoing work by various groups promises that the modified Raychaudhuri equation for the quadratic repulsive case can be indirectly tested using CMB observations in the near future  by analyzing the cosmological perturbations in loop quantum cosmology \cite{as}. Similarly, various works in brane world scenarios, which yield dynamical equations bearing similarity to the quadratic attractive case, have constrained parameters with supernovae and CMB experiments  (see for eg. \cite{roy,brane1}).  It can be hoped that future astronomical experiments might constrain the modification to Raychaudhuri equation, and lead us to the Hamiltonian for the modified gravity describing 
our universe at very large curvature scales. This is because our simple requirements prove so restrictive that they enable us to clinch the structure of 
canonical gravitational Hamiltonian directly from the properties of gravity encoded in dynamical equations.

%In a similar way, our simple requirements prove so restrictive that they enable us to clinch the structure of 
%canonical gravitational Hamiltonian directly from the properties of gravity encoded in dynamical equations. 
%Ongoing work by various groups promises that the modified Raychaudhuri equation for the quadratic repulsive case can be indirectly tested using CMB observations by analyzing the cosmological perturbations in loop quantum cosmology \cite{as}. Similarly, various works in brane world scenarios, which yield dynamical equations bearing similarity to the quadratic attractive case, have constrained parameters with supernovae and CMB experiments  (see for eg. \cite{roy,brane1}). It can be hoped that future precise astronomical experiments might constrain the modification to Raychaudhuri equation, and lead us to the Hamiltonian for the modified gravity describing our universe at very large curvature scales.

What is surprising is that just by reversing the canonical principle we can shed powerful light on the connection between gravitational phase space 
of the underlying theory and the repulsive character of modified gravity above a critical scale. Our 
findings reveal connections between mentioned modified gravity scenarios where degrees of freedom do not change and the structure of the gravitational phase space: No repulsive 
gravity? No polymerization!

\section*{Acknowledgments}

\noindent 
We are grateful to an anonymous referee for valuable suggestions which increased the clarity of our manuscript. 
This work is supported by NSF grants PHY-1403943 and PHY-1454832.

\end{document}